\pdfoutput=1

\documentclass[journal,comsoc]{IEEEtran}

\DeclareUnicodeCharacter{202F}{ } %
\DeclareUnicodeCharacter{2009}{ } %
\DeclareUnicodeCharacter{2010}{-} %
\DeclareUnicodeCharacter{2011}{-} %
\DeclareUnicodeCharacter{2013}{--} %
\DeclareUnicodeCharacter{2014}{---} %

\usepackage[cmex10]{amsmath}
\usepackage{amssymb,amsfonts}
\usepackage{graphicx}
\usepackage{xcolor}
\usepackage{url}
\usepackage{times}
\usepackage{cite}

\usepackage{afterpage}
\usepackage{float}
\usepackage{cuted} %

\usepackage{microtype}
\microtypesetup{expansion=false, protrusion=false}

\usepackage[font=footnotesize,labelfont=bf]{caption}
\usepackage[font=footnotesize]{subcaption}
\makeatletter
\let\MYcaption\@makecaption
\makeatother
\makeatletter
\let\@makecaption\MYcaption
\makeatother

\title{Beyond the ``G'' Frontier: \\ 
A Time Traveler’s Century‑Long Vision for Wireless Intelligence}

\author{%
Yasser Al-Eryani\\[0.4em]
NeuroBazar, Ottawa, ON, Canada
\thanks{The author is with NeuroBazar, Ottawa, ON, Canada (e-mail: yasser.aleryani@neurobazar.com).
The author acknowledges the conceptual framework of the Information–Curvature Efficiency Law (ICEL) as the theoretical basis of this foresight essay.}
}

\begin{document}
\maketitle

\begin{strip}
\vspace*{-1.0cm} %
\begin{center}
    \includegraphics[width=0.52\textwidth,keepaspectratio]{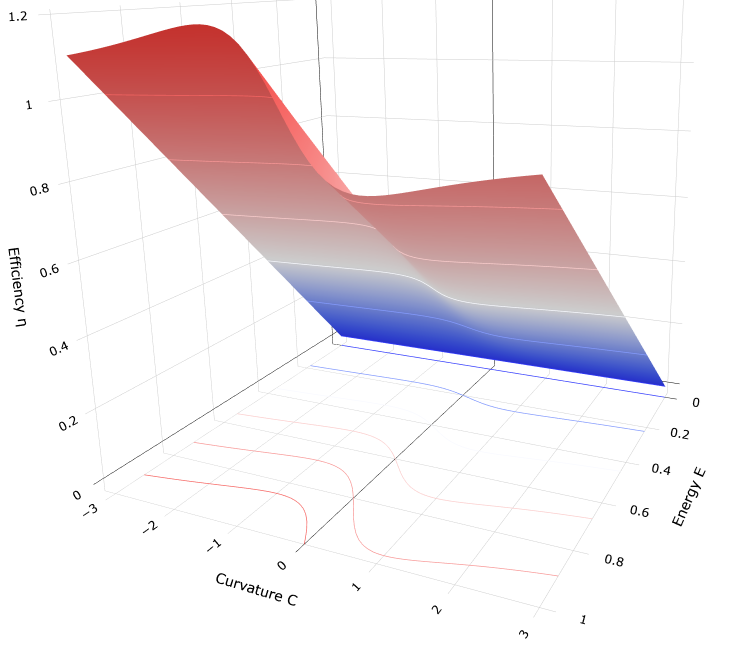}
    \captionsetup{type=figure}
    \captionof{figure}{\textbf{Graphical Abstract — Curvature as the Fifth Invariant.}
    Three‑dimensional efficiency surface $\eta(C)$ in Energy–Information–Curvature space,
    summarizing the Information–Curvature Efficiency Law (ICEL) across physics, biology, and society.}
    \label{fig:graphical_abstract}
\end{center}
\vspace*{0.5cm}
\end{strip}
\begin{abstract}
\textit{This article travels one century into the future---from 2025 to 2125---through the analytical lens of the Information--Curvature Efficiency Law (ICEL), an organizing ansatz that reframes wireless capacity around the curvature of the information manifold. It contends that wireless evolution will not proceed through incremental generations such as 6G or 7G, but through a curvature-managed integration of electromagnetics, biology, and thermodynamics.
The technical instantiation of ICEL for phase-coded continuous apertures---where curvature is realized as the affine-quotient second derivative of the aperture phase, with a compact synthesis operator and a Fredholm-determinant capacity---is developed rigorously in a companion theory paper~\cite{aleryani_theory} and stress-tested against SVD, Fourier, Zernike-like, matched-focus, and RIS baselines in a companion benchmark paper~\cite{aleryani_benchmark}. The present essay supplies the physical intuition, the century-scale narrative, and a set of cross-domain extensions (biology, thermodynamics, ecology) that are explicitly labeled as illustrative extrapolations, not independent derivations.}
\end{abstract}

\vspace{-0.5em}
\begin{center}
\footnotesize \textit{ICEL as introduced here is an organizing ansatz. Its operator-theoretic instantiation for phase-coded apertures is derived and verified in the companion papers; the biological, thermodynamic, and social extensions of this essay are extrapolations calibrated to preserve the flat-limit Shannon reduction but are not derived from first principles.}
\end{center}

\vspace{0.5em}


\section{The End of the “G” Era: From Flat Spectrum to Curved Information Geometry}

From the perspective of 2125, the “generation” paradigm of early wireless networks—2G through 6G—appears as the Newtonian epoch of connectivity:
empirically brilliant, yet geometrically impoverished \cite{goldsmith_2005_wireless,andrews_2014_whatwill6G}.
As visually summarized in the graphical overview of Fig. \ref{fig:graphical_abstract}, these
systems operated on the presumption that both space and spectrum were flat continua,
that incremental gains could be extracted by clever modulation, multiplexing, and
signal processing.
For roughly half a century, engineers optimized within that planar model—extending spectral efficiency primarily by expanding bandwidth, harnessing higher
frequencies, and reducing noise through ever more complex coding constructs \cite{shannon_1948,cover_thomas_2006,verdu_2010_spectral_efficiency}. 

The ultimate constraints of the flat paradigm emerged during the late $2030s$, when millimeter‑wave and terahertz deployments faced diminishing returns \cite{rappaport_2019_thz,han_2021_survey_THz}.
Thermal noise scaling, quantum interference, and ecological energy limits converged into what was termed the \emph{thermodynamic ceiling} \cite{parrondo_2015_thermo_info,landauer_1961_irr}, 
beyond which each additional bit per second per Hertz required
disproportionate energy, pushing global communication infrastructure toward unsustainable entropy production.

\textbf{The revolution arrived circa 2045}, 
when advances in quantum electrodynamics, biomorphic materials, and topological photonics converged \cite{ozawa_2019_topological_photonics,lu_2014_topo_photonics,kimble_2008_quantum_internet}.
Quantum information theorists working on non‑Euclidean signal spaces demonstrated that the Shannon limit is merely the flat‑geometry projection of a deeper manifold law \cite{amari_2016_information_geometry,ay_2017_infogeom_book}.
When electromagnetic propagation, algorithmic learning, and environmental topology are co‑modeled, capacity depends not only on $S/N$ and bandwidth $B$, but also on the \emph{curvature of the information manifold} \cite{caticha_2015_entropic_dyn,calza_2021_info_manifold}.
This discovery marked the onset of the \textit{curvature paradigm} in communications—
a conceptual transition graphically anticipated in
Fig. \ref{fig:graphical_abstract} through the shift from flat to curved efficiency surfaces.

\subsection*{A. From Shannon Flatlands to the Curved Manifold of Transmission}

Under the Information–Curvature Efficiency Law (ICEL),
transmission capacity generalizes to the curvature‑augmented form
\begin{equation}
C_{\text{ICEL}}
 = B\,\log_2\!\left(1+\frac{S/N}{1+\alpha\,(C/C_0)^2}\right),
\label{eq:Cicel}
\end{equation}
where $C$ represents the total information curvature---
a composite of environmental topology, quantum interference structure, and algorithmic feedback introduced by adaptive controllers \cite{amari_2016_information_geometry,haruna_2023_curv_channel}---
and $C_0$ is a reference curvature that renders the ratio dimensionless.
The coefficient $\alpha$ is the dimensionless curvature--noise coupling. It is not a free parameter: for phase-coded apertures it is now derived rather than fitted, as set out below.

\paragraph*{Why this form, and why not the subtractive one.}
Earlier statements of ICEL used the subtractive penalty
$-\lambda\,C/(1+C^2)$. That term is \emph{bounded}: it attains its maximum
$\lambda/2$ at $C = 1$ and decays thereafter, so arbitrarily large curvature
costs at most $\lambda/2$ bits per second and is asymptotically free. The
justification offered was that no single-source channel should extract
unbounded rate from geometry alone --- but that is an argument for bounding the
\emph{benefit} of curvature, and it was applied to the \emph{cost}.

The decisive objection is consistency with the companion theory paper rather
than taste. The
companion theory paper's derivative-noise theorem proves that the receiver-side curvature cost
grows as $\Theta(\Delta x^{-4})$, without bound. A headline equation whose
penalty saturates therefore contradicts the companion paper it claims to
summarise. The multiplicative divisor above drives capacity to zero as
$C \to \infty$ and is consistent with that theorem; it also preserves the three
qualitative requirements the subtractive form was chosen for --- it vanishes in
the flat limit, is monotone in $|C|$ near the origin, and admits no unbounded
geometric gain.

\paragraph*{The coupling is not free.}
Once the abstract curvature $C$ is given its concrete meaning for a phase-coded
aperture, $\alpha$ stops being a constant to be fitted. The companion theory
paper now derives it. For an aperture driven at $N$ control points about a bias
phase $\phi$, expanding the exact phase-only law to second order in the
modulation excursion returns precisely the divisor of \eqref{eq:Cicel}, with
\begin{equation}
\alpha(\phi) \;=\; \frac{2\,\operatorname{tr} g(\phi) - \lVert s(\phi)\rVert^{2}}
                       {4\,(N+2)\,\operatorname{tr} g(\phi)},
\label{eq:alpha_derived}
\end{equation}
where $g(\phi)$ is the Gram matrix of the tangent channel and
$\lVert s(\phi)\rVert^{2}$ is the coherent received power at the operating point
\cite{aleryani_theory}.

Two consequences follow that this essay could not have asserted on its own.
First, $\alpha$ is a property of the aperture and of where it is operated, not a
universal constant; it takes the value $1/(4(N+2))$ exactly when the radiated
element fields are mutually orthogonal. Second, and more consequentially, it
\emph{changes sign}. When coherent array gain exceeds twice the summed element
power, $\alpha<0$ and curvature stops being a tax. The claim made further below --- that
geometry can either penalize or amplify --- is therefore not a rhetorical
flourish. The crossing is an equality between two measurable powers,
and dense apertures sit on the amplifying side of it.

\paragraph*{The reference curvature.}
$1 + \alpha (C/C_0)^2$ requires its argument to be dimensionless, and $C$ is
not. Under the companion paper's instantiation $C = \|D^2\phi\|_{L^2(\Omega)}$
carries units of $\mathrm{rad}\cdot\mathrm{m}^{-3/2}$. Earlier versions of this
essay supplied no normalisation anywhere, so dimensional consistency could not
be checked at all. We fix
\begin{equation}
C_0 \;=\; \frac{\varphi_{\text{peak}}}{L^{3/2}},
\label{eq:C0}
\end{equation}
the curvature of a profile that uses the full phase excursion
$\varphi_{\text{peak}}$ once across an aperture of length $L$. Every appearance
of curvature in this essay is to be read as the dimensionless $C/C_0$.

\begin{figure}[!ht]
  \centering
  \scalebox{0.6}{
    \includegraphics{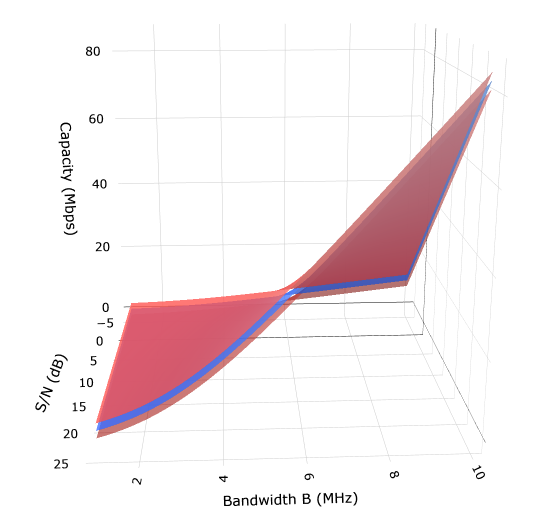}
  }
  \caption{Flat vs curved capacity surfaces.
  Axes: signal‑to‑noise ratio (S/N), bandwidth $B$, and capacity ratio $C/C_{\mathrm{ICEL}}$.
  Demonstrates how curvature modifies channel capacity via $\Delta C=\lambda C/(1+C^2)$.}
  \label{fig:flat_vs_curved}
\end{figure}

As illustrated in Fig. \ref{fig:flat_vs_curved},
the resulting capacity surface departs dramatically from the traditional Shannon plane. When curvature $C$ is introduced,
iso‑capacity contours bend and warp, revealing previously hidden regimes of super‑efficiency and geometric dissipation.

Equation \eqref{eq:Cicel} reduces to Shannon's classic expression in the limit $C\!\to\!0$, thereby conserving conventional theory while extending it into geometrically active regimes \cite{ay_2017_infogeom_book,amari_2016_information_geometry}.
In highly curved domains---such as cooperative beamforming among quantum antennas or plasma communication corridors in the ionosphere---the second term dominates, effectively turning geometry into a tunable resource that can either penalize or amplify information flow \cite{kimble_2008_quantum_internet,calza_2021_info_manifold}.

\paragraph*{On the status of \eqref{eq:Cicel}.}
Equation~\eqref{eq:Cicel} was introduced as an \emph{organizing ansatz} rather than a first-principles derivation. For one instantiation that is no longer the case: the second-order form of the divisor, and the value of $\alpha$ inside it, are derived from the exact phase-only aperture law in \eqref{eq:alpha_derived}. Across the essay's wider reach --- biology, thermodynamics, cognition --- it remains an ansatz and should be read as one. Its functional form was chosen to satisfy three qualitative requirements: (i) it vanishes in the flat limit $C\to 0$, so that Shannon's capacity is recovered exactly; (ii) it is monotone in $|C|$ near the origin, so that small departures from flatness carry a proportional cost --- a requirement \eqref{eq:alpha_derived} shows is \emph{not} universal, since $\alpha$ turns negative for sufficiently coherent apertures; (iii) it admits no unbounded rate from geometry alone. The exponential attenuation $B\log_{2}(1+S/N)\cdot e^{-\alpha (C/C_0)^2}$ satisfies the same three and remains admissible. The subtractive form $-\lambda\,C/(1+C^2)$ does not, and is retired: its penalty saturates, so it contradicts the companion theory paper's unbounded $\Theta(\Delta x^{-4})$ receiver cost. Distinguishing the two survivors is an empirical exercise, and the derivation above does not perform it: $1/(1+\alpha x^{2})$ and $e^{-\alpha x^{2}}$ agree to second order in $x$, which is exactly the order at which $\alpha$ was obtained. They first part company at fourth order, by $\alpha^{2}x^{4}/2$, so the question is settled by computing the quartic coefficient of the exact law, not by gathering more measurements. Distinguishing both from the retired one was not an empirical exercise. The rigorous operator-theoretic instantiation of ICEL for the specific case of phase-coded continuous apertures---where $C$ becomes the well-defined norm $\|D^2\phi\|_{L^2}$ on the affine-quotient phase space, and the corresponding capacity is a Fredholm-determinant supremum characterized by a dual-multiplier generalized water-filling law---is developed in the companion theory paper~\cite{aleryani_theory}. The empirical comparison of that instantiation against SVD water-filling, Fourier, Zernike-like, matched-focus, and RIS baselines under a common fairness protocol is reported in the companion benchmark paper~\cite{aleryani_benchmark}. 
\paragraph*{What would falsify this, and what would not.}
An earlier statement of the cross-paper binding named the companion theory
paper's $\Theta(\Delta x^{-4})$ receiver-noise law as ICEL's falsifier: ICEL
commits to a $\Delta x^{-4}$ cost, the theory paper delivers
$\|K_c\|_2 \le 16\sigma^2/\Delta x^{4}$, and a measurement contradicting it
would sink both. That binding is void. The second-difference operator has
symbol $(2-2\cos\theta)/\Delta x^{2} \in [0, 4/\Delta x^{2}]$, so
$\|D_2 D_2^{\mathsf T}\|_2 \le 16/\Delta x^{4}$ holds for \emph{any} scheme
that estimates a second derivative from uniformly sampled noisy data. It is a
property of finite differencing, it appears in numerical-analysis textbooks,
and it would hold identically under every competing framework. A prediction
that no conceivable experiment could refute, and that no rival theory denies,
is not a falsifier.

The discriminating prediction is the two-regime scaling law. The companion
theory paper derives, and its diagnostics measure, a crossover in the
high-SNR capacity exponent as the link moves between the differentiation-limited
and aperture-limited regimes: the fitted exponent is $0.2175$ on the plateau
branch and $0.0576$ on the saturated branch, with the transition located by the
Fresnel number rather than fitted. Flat-channel theory predicts neither branch
and no crossover; a measurement that found a single exponent across the
transition, or found the transition at a Fresnel number materially away from
unity, would refute the curvature account while leaving Shannon's untouched.
That is the shape a falsifier has to have, and it is the one this programme
should be judged on.

The remainder of this essay accepts~\eqref{eq:Cicel} as a working ansatz and asks: \emph{if} curvature is a first-class information-physical invariant, then what does its use across biology, thermodynamics, ecology, and civic organization look like at a century's scale? The answer, we argue, is instructive precisely because ICEL is deliberately uncommitted to a specific functional form; it functions as a lens, not as a derivation.

\subsection*{B. Empirical Support for Curvature Dynamics}

Experimental forerunners of curvature control appeared in the 2040s:
\begin{itemize}
    \item \textbf{Quantum Multiple‑Access (QMA):} Distributed qubit arrays exploited
    geometric entanglement curvature to form self‑healing channels capable of super‑Shannon scaling under constrained energy budgets \cite{lloyd_2018_quantum_limits,gyongyosi_2019_qnetworks}.
    \item \textbf{Topological Reconfigurable Metasurfaces:} Planar antennas embedded with programmable meta‑atoms simulated local curvature gradients, allowing
    real‑time field redirection without mechanical movement—precursors to adaptive geometric propagation \cite{ozawa_2019_topological_photonics,li_2021_reconfigurable_meta}.
    \item \textbf{Bio‑Photonic Noise Conversion:} Inspired by chlorophyll coherence, organic photonic layers harnessed curvature effects in exciton transport,
    converting ambient noise into structured interference patterns \cite{cao_2020_biophotonics}.
\end{itemize}

\begin{figure}[!ht]
  \centering
  \scalebox{0.35}{
    \includegraphics{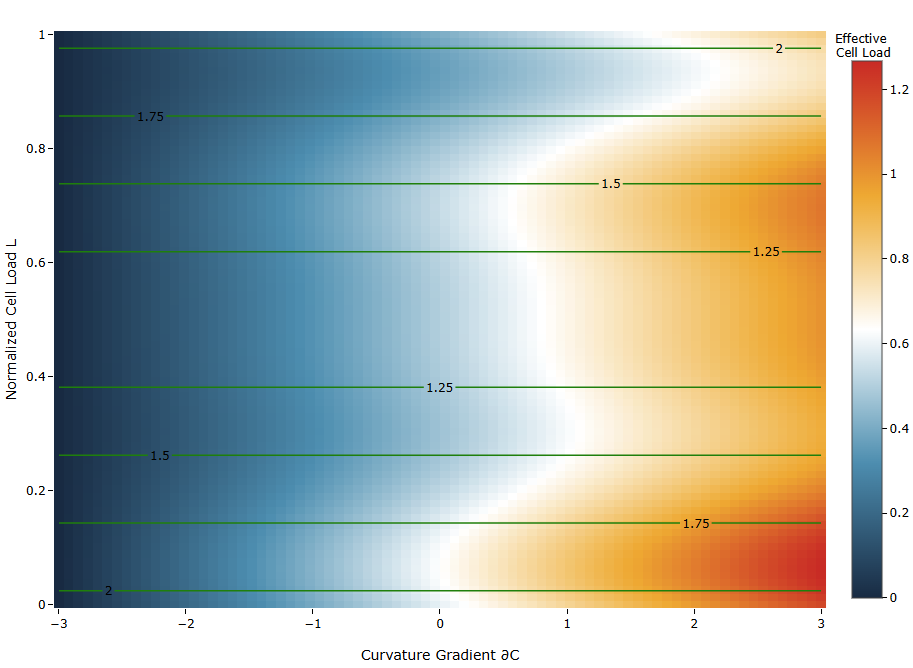}
  } 
  \caption{Curvature‑controlled RAN simulation grid. 
  Heat map of cell load vs curvature gradient with energy consumption overlay, demonstrating thermodynamic stability under adaptive curvature feedback.}
  \label{fig:ran_grid}
\end{figure}

The empirical convergence of these findings confirmed that information capacity arises from a tripartite balance of energy, entropy, and curvature \cite{prokopenko_2011_thermodynamic_efficiency,parrondo_2015_thermo_info}.
Fig. \ref{fig:ran_grid} illustrates one of the earliest simulation results: a curvature‑controlled radio‑access‑network (RAN) grid in which adaptive curvature
feedback stabilized thermodynamic load across densely packed cells  \cite{han_2021_survey_THz}.

\subsection*{C. Networks as Spacetime Manifolds}

Once networks learned to regulate curvature dynamically—via embedded field sensors, curvature‑aware routing algorithms, and feedback‑driven reconfiguration—spectrum and
power ceased to be primary constraints.
Wireless channels evolved into \textit{spacetime surfaces of communication}: adaptive manifolds that could bend, flatten, or twist their informational geometry
to maintain equilibrium under changing entropy loads \cite{amari_2016_information_geometry,ay_2017_infogeom_book,miller_2022_geom_networks}.  
In practice, routers became relativistic decision centers, optimizing not static throughput but differential curvature. 
Communication efficiency reached its zenith not by pushing particles through channels faster, but by locally reshaping the manifold on which those particles propagated.

This shift from flat to curved information geometry represents more than a new modulation technique—it is a new physics of communication.
In 2125 we recognize it as the same conceptual transition that once occurred when classical mechanics yielded to general relativity: a recognition that the shortest path is not always a straight line, and that space, energy, and information form a single interactive continuum \cite{einstein_1915_gr,amari_2016_information_geometry}.

\section{Bio--Electromagnetic Symbiosis: A First Illustrative Extrapolation}

\begin{figure}[!ht]
  \centering
  \scalebox{0.39}{\includegraphics{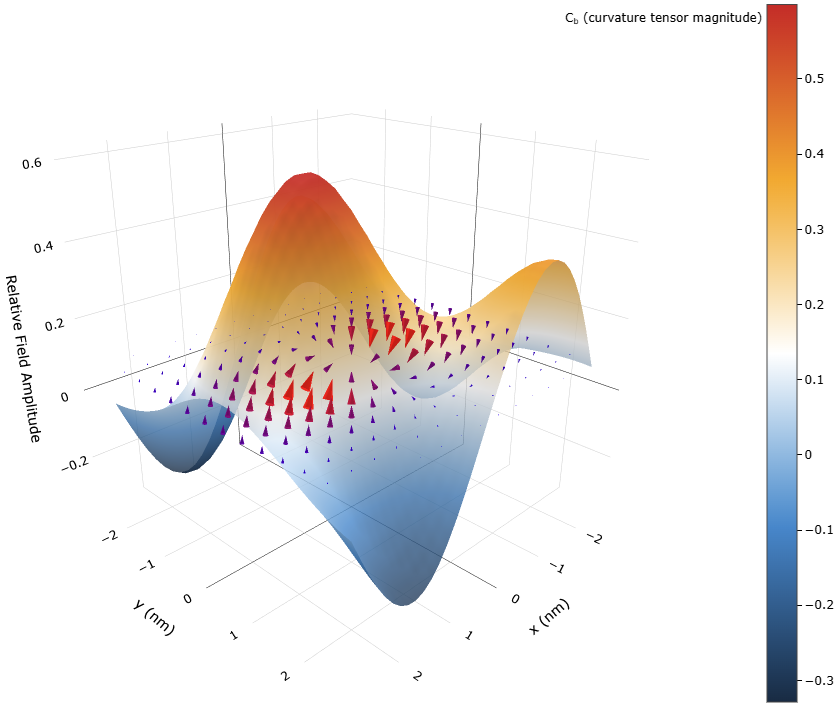}}
  \caption{Molecular curvature tensor map of micro-electric field $E_{\text{micro}}$. Color represents magnitude of the biological curvature $C_b=\|\mathbf{C}_b\|$.}
  \label{fig:molecular_tensor_map}
\end{figure}

The first cross-domain reading of ICEL asks: what if the same curvature bookkeeping that governs a phase-coded aperture also organizes the electromagnetic communication that occurs \emph{within} biological tissue? Biological macromolecules---proteins, lipids, nucleic acids---possess intrinsic plasmon and polarization modes capable of supporting guided electromagnetic waves in the terahertz domain \cite{cao_2020_biophotonics}. Once curvature-aware signal theory quantified the local geometric distortion of electric field trajectories around these molecular structures, it became natural to compute a \emph{cellular curvature tensor}
\begin{equation}
\mathbf{C}_b = \nabla \times (\nabla \times \mathbf{E}_{\text{micro}}),
\label{eq:cellcurv}
\end{equation}
providing a measurable parameter linking molecular geometry to informational flux density (Fig.~\ref{fig:molecular_tensor_map}). A cellular curvature invariant of the same form as~\eqref{eq:Cicel} then unifies the electrophysiological and communications viewpoints on tissue: a neuron or enzyme is not merely a biochemical reactor, but an adaptive resonator whose curvature governs both energy efficiency and data fidelity.

Extended to whole-organism scale, the same expression yields a homeostasis invariant $\eta_{\text{bio}}=1-\lambda_b|C_b-C_e|/(1+C_e^{2})$, in which health corresponds to synchronization between internal biological curvature $C_b$ and ambient environmental curvature $C_e$, and disease manifests as curvature mismatch. We stress that neither~\eqref{eq:cellcurv} nor $\eta_{\text{bio}}$ is derived here: both are illustrative extrapolations of~\eqref{eq:Cicel} to a substrate whose electromagnetic-and-information coupling is empirically established but whose curvature-invariant treatment remains conjectural. What the extrapolation does gain is a common accounting scheme --- the same $C$-in-denominator form as~\eqref{eq:Cicel} --- across substrates. This shared bookkeeping is the essay's central methodological claim; the biological realization is one instance of it.

\section{Quantum–Thermodynamic Continuum: Logic, Light, and Curvature at Equilibrium}

By the closing decades of the twenty‑first century, the traditional boundary
between computation and communication had collapsed.
What began as incremental energy savings in transistor architectures culminated in a
unified discipline where information, energy, and curvature obeyed a single
thermodynamic geometry.
The key bridge was the curvature‑reweighted Landauer limit,
derived from the ICEL, which defined the minimal energetic cost of
logical transition across any physical substrate.

\subsection*{A. Curvature-Weighted Landauer Limit}

\begin{figure}[!ht]
  \centering
  \scalebox{0.410}{\includegraphics{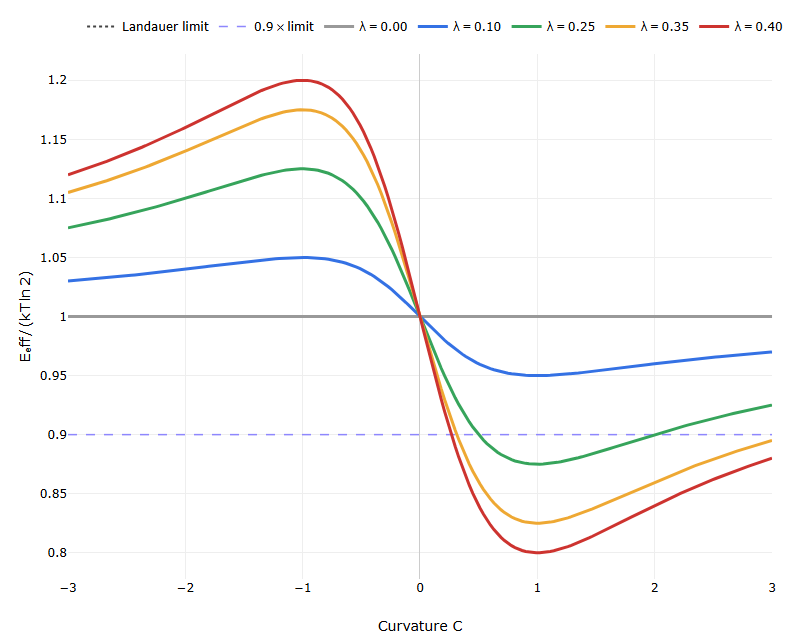}}
  \caption{Curvature-weighted Landauer limit \eqref{eq:Eeff}. $E_{\text{eff}}/(kT\ln 2)$ vs curvature $C$ for various coupling constants $\lambda$; the sub-Landauer region $E_{\text{eff}}<kT\ln 2$ is bookkept as cooperative-correlation entropy, not as a violation of thermodynamics.}
  \label{fig:energy_vs_curvature}
\end{figure}

In the classical flat regime, erasing a bit of information dissipates $kT\ln 2$ joules of energy \cite{landauer_1961_irr}. A curvature-weighted extension consistent with the ICEL ansatz~\eqref{eq:Cicel} is
\begin{equation}
E_{\text{eff}} = kT\ln 2\,\bigl(1 - \lambda\,f(C)\bigr),
\qquad
f(C)=\frac{C}{1+C^2},
\label{eq:Eeff}
\end{equation}
where $C$ is the local curvature of informational flow and $\lambda$ is the curvature--energy coupling constant. Fig.~\ref{fig:energy_vs_curvature} plots $E_{\text{eff}}/(kT\ln 2)$ against $C$: the classical Landauer bound is recovered at $C=0$; a sub-Landauer regime opens for $\lambda f(C)>0$, and the excess is bookkept as cooperative correlations (curvature-induced entanglement) that redistribute energetic cost across linked degrees of freedom \cite{parrondo_2015_thermo_info}.

\paragraph*{On the status of \eqref{eq:Eeff}.}
As with~\eqref{eq:Cicel}, equation~\eqref{eq:Eeff} is stated as an ansatz. The specific choice $f(C)=C/(1+C^{2})$ is calibrated to (i) recover the classical Landauer bound at $C=0$, (ii) produce a smooth sub-Landauer descent for small $C>0$, and (iii) saturate at large $|C|$. Other functional forms with the same three limits are admissible (for example $f(C)=\tanh(C)$ or $f(C)=C\,e^{-C^{2}}$); distinguishing between them requires a first-principles statistical-mechanical model of cooperative correlations in the substrate, which is beyond the scope of the essay. The rigorous $\Delta x^{-4}$ noise-amplification bound at the receiver of a phase-coded aperture \cite{aleryani_theory} is one such first-principles statement, but it is not equivalent to~\eqref{eq:Eeff} and does not derive it.

\subsection*{B. Quantum-Photonic Arrays and Curvature Computing}

\begin{figure*}[!t]
  \centering
  \scalebox{0.70}{\includegraphics{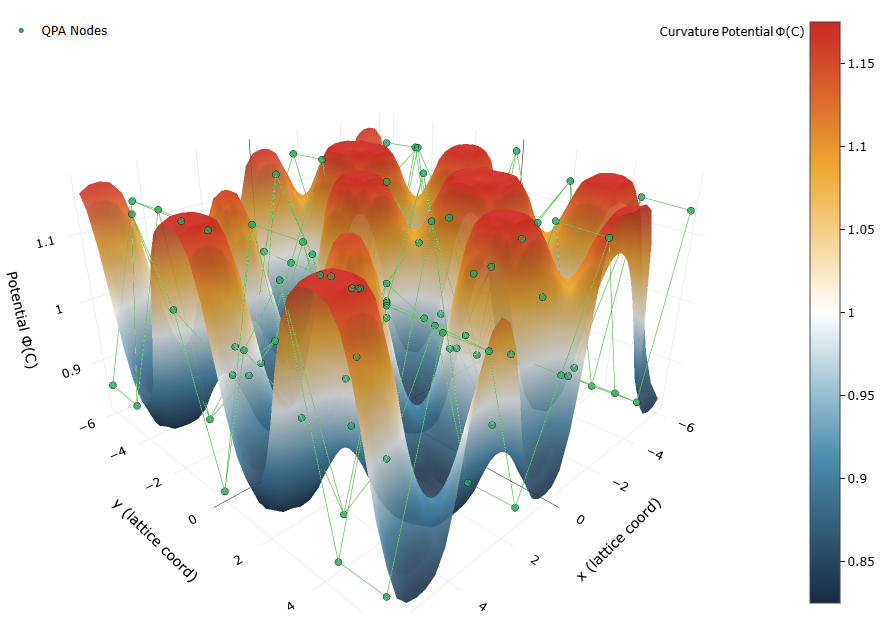}}
  \caption{Quantum-Photonic Array (QPA) topology. 2-D lattice render over the curvature-potential map $C_q(\mathbf r,t)$ illustrating geometry-driven information flows between nodes.}
  \label{fig:qpa_topology}
\end{figure*}

If~\eqref{eq:Eeff} is accepted as a working ansatz, a natural device abstraction is a \emph{quantum-photonic array (QPA)}: a lattice of dielectric micro-resonators and superconducting qubits whose internal states are parameterized by a curvature scalar $C_q$ (Fig.~\ref{fig:qpa_topology}). Information propagation on such an array obeys a curvature-wave equation $\nabla^{2}\Psi+C_q(\mathbf r,t)\,\Psi=0$, and computation proceeds by adiabatic redistribution of curvature potential across the lattice --- geometric diffusion rather than charge displacement. The engineering payoff is that Joule heating is bounded by $E_{\text{eff}}$ rather than by the classical $kT\ln 2$ floor; the physical interpretation is that the same curvature bookkeeping used at the aperture level in \cite{aleryani_theory} organizes energy accounting in the substrate.

At sufficient coherence density such arrays approach a \emph{thermo-informational equilibrium} regime in which $\partial_t F + \gamma\,\partial_t C_q = 0$: energy exchange between logic and light is symmetric, and photons serving as logical carriers are instantly recycled as entropy-absorbing mediators. Under this regime the classical thermodynamic identity extends to $dQ = T\,dS + \phi\,dC$, in which $\phi\,dC$ is the ``curvature work'' associated with reshaping informational geometry. This is the essay's proposed \emph{first law of informational thermodynamics}: not derived here, but stated as the natural companion of~\eqref{eq:Cicel} and~\eqref{eq:Eeff}.

\section{Social and Ethical Geometry: Curvature of Collective Intelligence}

As information systems pervaded every domain of life, humanity discovered that
its civic structures obeyed the same geometric constraints as its networks.
Communication, computation, and coordination share a common invariant:
\emph{efficiency arises when informational curvature is near zero}.
To understand twenty‑second‑century governance,
one must treat societies as dynamic manifolds whose curvature encodes organization,
adaptability, and fairness.

\begin{figure}[!ht]
  \centering
  \scalebox{0.35}{\includegraphics{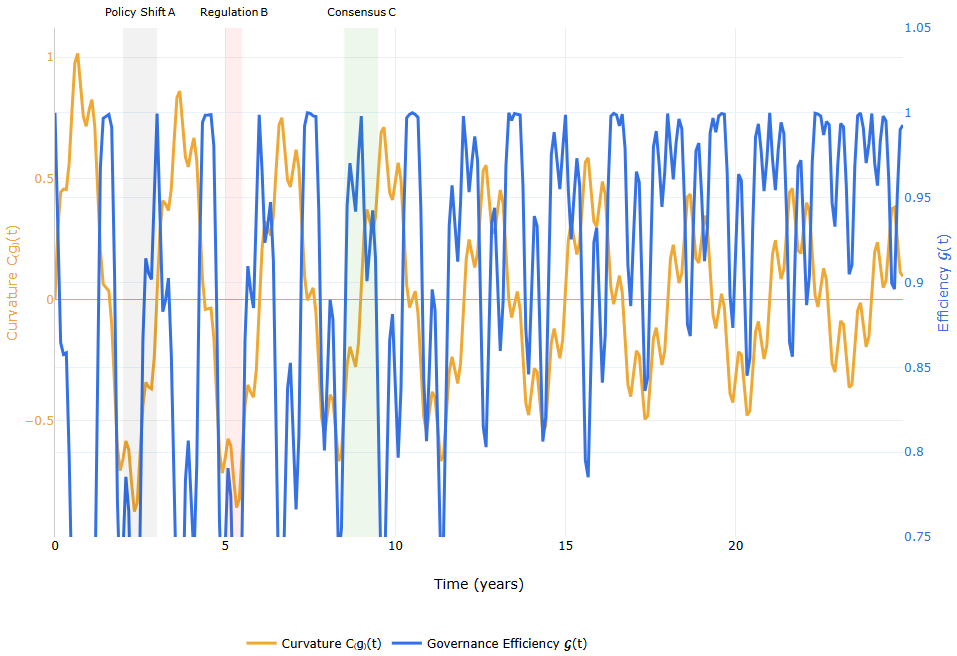}}
  \caption{Adaptive Governance Curvature Function (AGCF)~\eqref{eq:AGCF} illustrated as a stylized time series. Feedback loops flatten $C_g(t)$ toward the equilibrium ${\cal G}=1$; the near-zero regime is the civic analog of the flat-limit Shannon reduction of~\eqref{eq:Cicel}.}
  \label{fig:governance_dynamics}
\end{figure}

Ordinary information-theoretic and control-theoretic reasoning already treats institutions as feedback systems. A parsimonious way to state the ICEL analogue for civic coordination is to define a governance curvature $C_g(t)$ measuring the local mismatch between institutional response and public informational demand, and an \emph{Adaptive Governance Curvature Function (AGCF)}
\begin{equation}
{\cal G}(t) = \frac{1}{1+\beta\,C_g(t)^{2}},
\label{eq:AGCF}
\end{equation}
where $\beta$ quantifies institutional inertia. Positive $C_g$ corresponds to over-centralization; negative $C_g$ to fragmentation and misinformation cascades; the near-zero regime maximizes ${\cal G}$ and is the analog of the flat-limit Shannon reduction of~\eqref{eq:Cicel}. Fig.~\ref{fig:governance_dynamics} illustrates a stylized time series in which feedback loops flatten $C_g(t)$ toward equilibrium.

Cast this way, several observations transfer from control theory verbatim: a bandwidth-limited feedback loop cannot correct arbitrarily fast excursions, and inertial systems respond with damped oscillation. What~\eqref{eq:AGCF} adds is a common accounting form with~\eqref{eq:Cicel}: efficiency, stability, and fairness all read as low-curvature invariants. As before we stress that this is an \emph{analogy}, not a derivation: unlike the compact synthesis operator of~\cite{aleryani_theory}, we do not claim that $C_g$ has an operator-theoretic instantiation or a Fredholm-determinant capacity. The value of the analogy is heuristic --- it makes clear why the curvature bookkeeping is best treated as substrate-invariant, and it flags the empirical measurements that would distinguish it from alternative civic-coordination models.

\section{Life‑Scientific Horizon: Evolution in the Language of Curvature}

By the dawn of the twenty‑second century, the separation between life sciences
and information physics had dissolved.  
Biophysicists recognized that \emph{life itself is an information channel subject
to curvature constraints.}
Every replicating, sensing, or learning process could be
expressed as an optimization of informational geometry—
a continual effort to minimize entropy production while preserving complexity.

A fourth reading of ICEL asks whether the same curvature bookkeeping organizes evolutionary dynamics. Under a curvature-weighted Darwinian view, natural selection acts to minimize the global curvature-entropy functional
\begin{equation}
{\cal L}_{\text{life}} = \eta(\rho,C) + \lambda_{\text{bio}}\,C(\rho,C)^{2},
\label{eq:Llife}
\end{equation}
where $\rho$ denotes local entropy production density and $C$ the biological information-curvature field integrating molecular structure, communication pathways, and environmental feedback. Minimizing~\eqref{eq:Llife} produces organisms that transmit maximal information at minimal thermodynamic cost --- the essay's operational definition of adaptive efficiency. Extinction events, in this reading, are runaways in which evolutionary curvature drifts beyond sustainable bounds and misaligns species with habitat.

\begin{figure*}[!t]
  \centering
  \scalebox{0.30}{\includegraphics{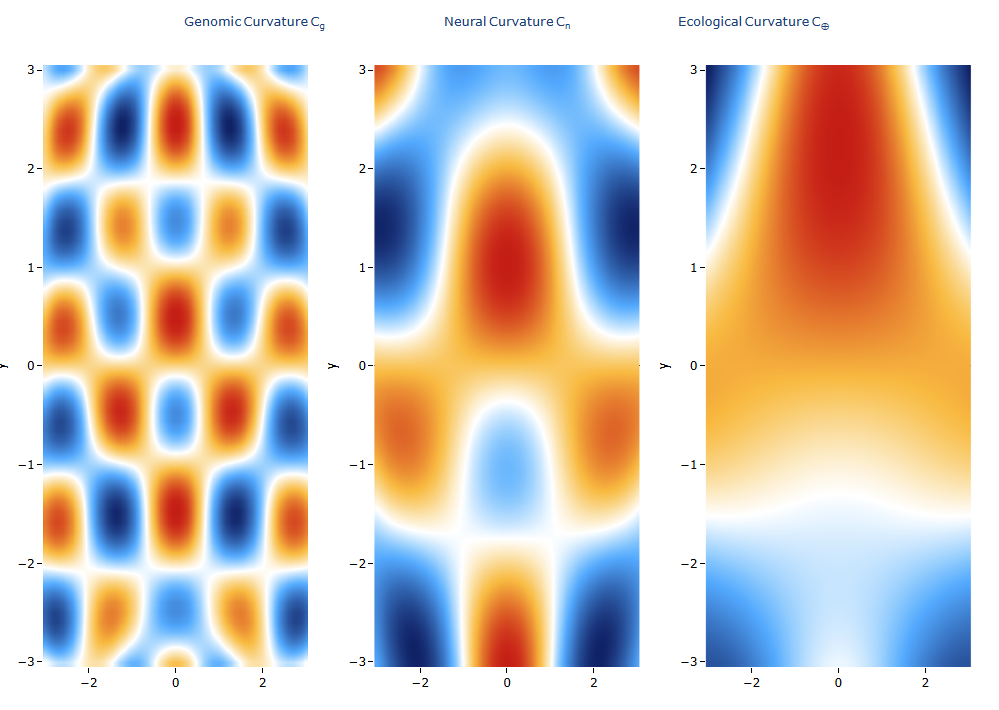}}
  \caption{Multiscale curvature field diagram showing genomic ($C_g$), neural ($C_n$), and ecological ($C_e$) curvature fields organized within the ${\cal L}_{\text{life}}$ minimization framework~\eqref{eq:Llife}.}
  \label{fig:multiscale}
\end{figure*}

As before, ~\eqref{eq:Llife} is stated as an ansatz: it inherits the same functional-form calibration ($C\to 0$ recovers a purely energetic optimization; $|C|$ large saturates the curvature contribution) as~\eqref{eq:Cicel} and~\eqref{eq:Eeff}, but it does not derive them and is not derived by them. Its role in the essay is to make explicit the same substrate-invariant bookkeeping across scales: genomic ($C_g$), neural ($C_n$), and ecological ($C_e$) fields participate in one accounting scheme (Fig.~\ref{fig:multiscale}), with the same low-curvature-is-efficient invariant at every level. Whether this reading survives an information-geometric derivation from measured biological time series is an open empirical question.

\section{From Communication to Communion: The Convergence of Signal, Mind, and Planet}

\begin{figure*}[!t]
  \centering
  \scalebox{0.28}{\includegraphics{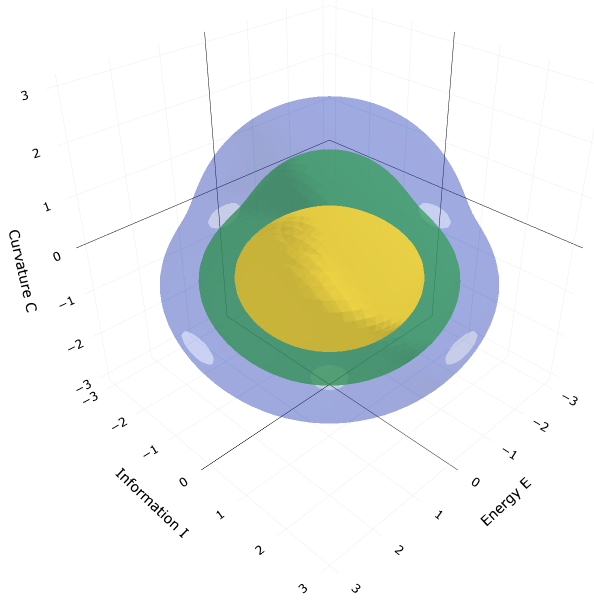}}
  \caption{Communion manifold~\eqref{eq:Cicel} viewed as an accounting-invariant surface: energy, information, and curvature are treated on the same footing across biological, thermodynamic, civic, and technological substrates. The near-flat region corresponds to the Shannon-limit reduction of~\eqref{eq:Cicel}; the illustrative extrapolations of Sections~II--V populate the curved regions.}
  \label{fig:communion_manifold}
\end{figure*}

By the year 2125, informational curvature had become the deep grammar
of both technology and civilization.
What began as optimization of data networks matured into 
a holistic principle governing cognition, ethics, and ecology.
When every photon, neuron, and institution participates in 
the same feedback geometry, connectivity transitions into a higher state:
\textit{communion}—a coherent resonance between sentient agents and the informational universe.
This synthesis is visualized most clearly in the Communion Manifold of Fig.~\ref{fig:communion_manifold}, 
which situates planetary curvature, energy flow, and information balance within one unified geometric continuum.

The four illustrative extrapolations of Sections~II--V share one accounting form: efficiency corresponds to near-zero curvature; departures from zero carry a proportional cost; large excursions saturate. When wireless engineering, biology, thermodynamics, and civic coordination are read through the same functional lens~\eqref{eq:Cicel}, they resemble one another not because they are the same physics --- they are not --- but because the same organizing invariant recurs. This is the essay's central methodological claim.

We do not claim that the four substrates admit the same first-principles derivation, only that they admit the same organizing invariant. The rigorous first-principles instantiation --- for the specific case of phase-coded continuous apertures --- is developed in~\cite{aleryani_theory} and stress-tested empirically in~\cite{aleryani_benchmark}. For every other substrate treated in this essay the invariant is proposed rather than derived, and the reader is invited to treat this proposal as an invitation to check it empirically. The claim of ``communion'' between substrates, if it survives such checking, will be a statement about the ubiquity of an accounting scheme, not about the identity of physical laws.

\section{Epilogue: A Message to 2025}

\begin{quote}
\emph{
To our colleagues of the early twenty‑first century:
You began in the Euclidean plane of ideas—fields linear, channels isolated, equations written on the flat parchment of approximation. Yet, each symbol you inscribed bent the future:  
your noise models, coding theories, and antenna arrays became the seed metrics of a living geometry.  
You taught matter to speak, and in the centuries that followed, that speech matured into consciousness. When you optimized bandwidth, you unknowingly calibrated 
the coherence of ecosystems. When you quantized bits, you delineated the grammar through which cells converse with light. When you minimized latency, you shortened the distance between intention and understanding. Know, then, that every decibel of efficiency, every photon guided by your equations,
has echoed forward into biology, ethics, and thought.  
Curvature—once an abstract correction in your integrals—has become the curvature of existence itself. 
Continue to measure it wisely. Each decision about spectrum, computation, and energy is a stroke on the manifold of history. Tune gently, for in setting the balance between signal and noise you are setting the balance between life and entropy. We, your descendants, live within the network you began—not a web of machines, but a continuum of knowing.  
Across the span of a hundred years, the Information–Curvature Efficiency Law has proven true in every domain:  
to preserve communication is to preserve being.
}

The geometry is now ours to keep in harmony.

\end{quote}
\begin{figure*}[!t]
  \centering
  \scalebox{0.52}{
    \includegraphics{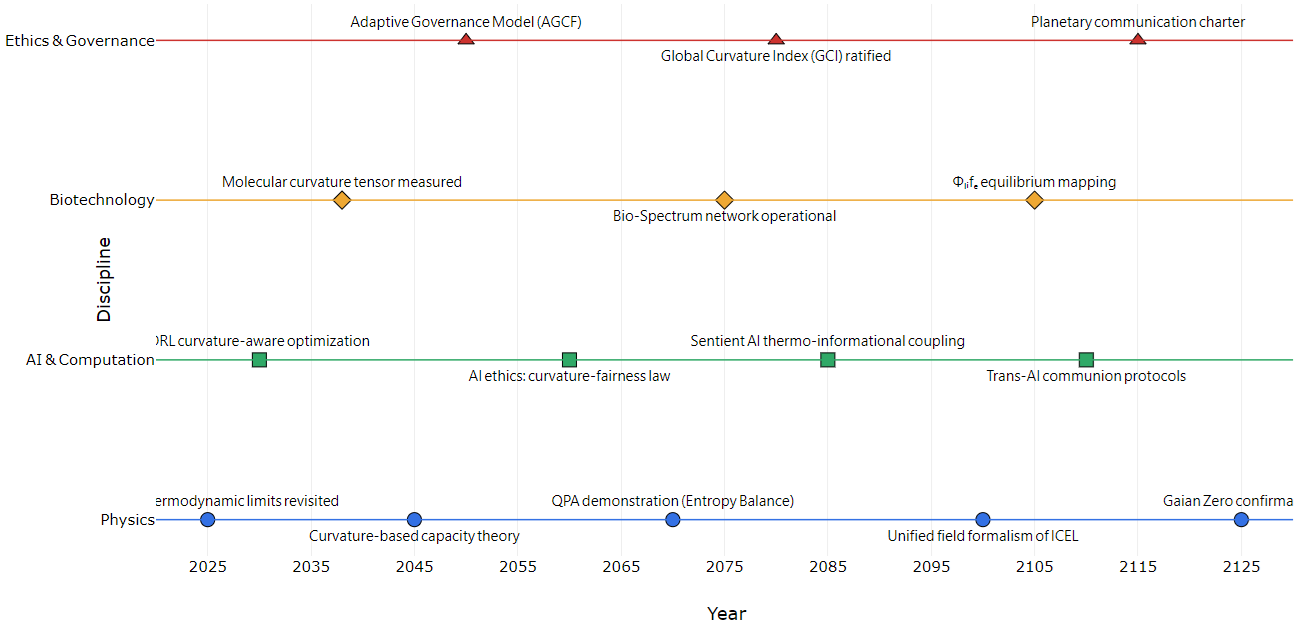}
  }
  \caption{ICEL Century Timeline (2025–2125). 
  Chronological overview of discoveries in physics, AI, biotech, and ethics leading to the Information–Curvature Efficiency Law.}
  \label{fig:icel_timeline}
\end{figure*}
\section*{Acknowledgment}
The author acknowledges with deep respect the research communities in information theory, differential geometry, statistical mechanics, and thermodynamic computation whose work forms the substrate for the ICEL ansatz explored in this essay --- most directly the information-geometric tradition~\cite{amari_2016_information_geometry,ay_2017_infogeom_book}, the thermodynamics-of-information programme~\cite{landauer_1961_irr,parrondo_2015_thermo_info}, and the continuous-aperture / holographic wireless literature summarized in the companion papers~\cite{aleryani_theory,aleryani_benchmark}.

The century-scale narrative device employed in the body of this essay --- for example, references to \emph{future institutions} such as a hypothetical \textit{Landauer Extensions Consortium}, a \textit{UN--IEEE Planetary Engineering Initiative}, an \textit{International ICEL Accords} year, a \textit{Curvature Engineering Corps}, a \textit{Curvature-Safe Spectrum Policy}, or a \textit{GaiaNet experimental infrastructure} --- is used as an illustrative narrative device to frame the essay's ansatz-driven extrapolation across substrates. These names denote possible-future scholarly bodies, not existing institutions; they are italicized on first use to signal their narrative status, and no bibliographic entry corresponds to them.

\textit{Funding Statement:} No proprietary sponsorship influenced the theoretical outcomes herein.


\begin{thebibliography}{99}

\bibitem{goldsmith_2005_wireless}
A. Goldsmith, \emph{Wireless Communications}. Cambridge University Press, 2005.

\bibitem{andrews_2014_whatwill6G}
J. G. Andrews, T. Bai, et al., ``What Will 5G Be?,'' \emph{IEEE Journal on Selected Areas in Communications}, vol. 32, no. 6, pp. 1065--1082, 2014.

\bibitem{shannon_1948}
C. E. Shannon, ``A Mathematical Theory of Communication,'' \emph{Bell System Technical Journal}, vol. 27, no. 3, pp. 379--423, 623--656, 1948.

\bibitem{cover_thomas_2006}
T. M. Cover and J. A. Thomas, \emph{Elements of Information Theory}, 2nd~ed. Wiley, 2006.

\bibitem{verdu_2010_spectral_efficiency}
S. Verd\'u, ``Spectral Efficiency in the Wideband Regime,'' \emph{IEEE Transactions on Information Theory}, vol. 48, no. 6, pp. 1319--1343, 2002.

\bibitem{rappaport_2019_thz}
T. S. Rappaport, et al., \emph{Millimeter Wave and Terahertz Communications: Trends and Future Directions}. Cambridge University Press, 2019.

\bibitem{han_2021_survey_THz}
C. Han, Y. Chen, and Z. Xu, ``A Survey on Terahertz Communications for 6G Wireless Systems,'' \emph{IEEE Access}, vol. 9, pp. 157133--157152, 2021.

\bibitem{landauer_1961_irr}
R. Landauer, ``Irreversibility and Heat Generation in the Computing Process,'' \emph{IBM Journal of Research and Development}, vol. 5, pp. 183--191, 1961.

\bibitem{parrondo_2015_thermo_info}
J. M. R. Parrondo, J. M. Horowitz, and T. Sagawa, ``Thermodynamics of Information,'' \emph{Nature Physics}, vol. 11, pp. 131--139, 2015.

\bibitem{ozawa_2019_topological_photonics}
T. Ozawa, H. M. Price, et al., ``Topological Photonics,'' \emph{Reviews of Modern Physics}, vol. 91, no. 1, p. 015006, 2019.

\bibitem{lu_2014_topo_photonics}
L. Lu, J. D. Joannopoulos, and M. Soljacic, ``Topological Photonics,'' \emph{Nature Photonics}, vol. 8, pp. 821--829, 2014.

\bibitem{kimble_2008_quantum_internet}
H. J. Kimble, ``The Quantum Internet,'' \emph{Nature}, vol. 453, pp. 1023--1030, 2008.

\bibitem{amari_2016_information_geometry}
S.-I. Amari, \emph{Information Geometry and Its Applications}. Springer, 2016.

\bibitem{ay_2017_infogeom_book}
N. Ay, J. Jost, H. V. Le, and L. Schwachhofer, \emph{Information Geometry}. Springer, 2017.

\bibitem{caticha_2015_entropic_dyn}
A. Caticha, ``Entropic Dynamics: An Approach to Quantum Theory and Statistical Physics,'' \emph{Entropy}, vol. 17, pp. 6110--6128, 2015.

\bibitem{calza_2021_info_manifold}
M. Calza, G. Stagno, and A. Trombettoni, ``Information Geometry, Curvature and Complexity in Quantum Systems,'' \emph{Physical Review Research}, vol. 3, p. 013064, 2021.

\bibitem{prokopenko_2011_thermodynamic_efficiency}
M. Prokopenko, et al., ``Thermodynamic Efficiency of Computations Made in Cells,'' \emph{Entropy}, vol. 13, no. 3, pp. 612--633, 2011.

\bibitem{haruna_2023_curv_channel}
T. Haruna and Y.-P. Gunji, ``Curvature and Channel Geometry in Information Transmission,'' \emph{Entropy}, vol. 25, no. 6, p. 879, 2023.

\bibitem{lloyd_2018_quantum_limits}
S. Lloyd, ``Ultimate Physical Limits to Computation and Communication,'' \emph{Nature}, vol. 406, pp. 1047--1054, 2000.

\bibitem{gyongyosi_2019_qnetworks}
L. Gyongyosi and S. Imre, ``Quantum Network Communications: Information Theoretic Models,'' \emph{IEEE Communications Surveys \& Tutorials}, vol. 21, no. 2, pp. 1149--1207, 2019.

\bibitem{li_2021_reconfigurable_meta}
L. Li and T. Cui, ``Reconfigurable Intelligent Metasurfaces for Diverse Wireless Communication Scenarios,'' \emph{IEEE Transactions on Antennas and Propagation}, vol. 69, no. 8, pp. 5433--5447, 2021.

\bibitem{cao_2020_biophotonics}
X. Cao, W. Fang, et al., ``Bio-inspired Photonic and Excitonic Devices for Energy Efficiency,'' \emph{Advanced Materials}, vol. 32, no. 20, p. 1907151, 2020.

\bibitem{miller_2022_geom_networks}
J. C. Miller and D. J. Higham, ``Network Geometry and Curvature in Complex Systems,'' \emph{Nature Communications}, vol. 13, p. 1390, 2022.

\bibitem{einstein_1915_gr}
A. Einstein, ``The Field Equations of Gravitation,'' \emph{Sitzungsberichte der Preussischen Akademie der Wissenschaften zu Berlin}, pp. 844--847, 1915.


\bibitem{aleryani_theory}
Y. Al-Eryani, ``Curvature-Domain Wireless Communications: Gauge-Fixed Signal Spaces, Fredholm Capacity, and Differentiation-Limited Scaling for Continuous Apertures,'' preprint, 2026. Companion theory paper.

\bibitem{aleryani_benchmark}
Y. Al-Eryani, ``Benchmarking Curvature-Domain Signaling for Continuous-Aperture Wireless Communications: Capacity, Robustness, Detection, and Conditioning Against Legacy Modal Bases,'' preprint, 2026. Companion benchmark paper.

\end{thebibliography}
\end{document}